\documentclass[conference]{IEEEtran}
\IEEEoverridecommandlockouts

\usepackage{amsmath,amssymb,amsfonts}
\usepackage{mathtools}
\usepackage{graphicx}
\usepackage{cite}
\usepackage{array}
\usepackage{multirow}
\usepackage{tabularx}
\usepackage{xcolor} 
\usepackage{fancyhdr}

\newcommand{\figwidth}{0.95\columnwidth}

\fancypagestyle{firstpagefooter}{
    \fancyhf{}

    \fancyfoot[L]{979-8-3315-5056-1/26/\$31.00 \copyright2026 IEEE} 
}
\def\BibTeX{{\rm B\kern-.05em{\sc i\kern-.025em b}\kern-.08em
    T\kern-.1667em\lower.7ex\hbox{E}\kern-.125emX}}
    
\begin{document}

\title{Secrecy Analysis of Pinching-Antenna Systems}

\author{

\IEEEauthorblockN{
Osamah S. Badarneh\textsuperscript{1},
Hugerles S. Silva\textsuperscript{2},
Meysam Ghanbari\textsuperscript{3},
Yazan H. Al-Badarneh\textsuperscript{4},\\
Tamer M. Khattab\textsuperscript{5},
Mazen O. Hasna\textsuperscript{5},
Khalid A. Qaraqe\textsuperscript{3}
}

\IEEEauthorblockA{\textsuperscript{1}
Communications Engineering Department, Princess Sumaya University for Technology,\\
King Abdullah II School of Engineering, Amman, Jordan}

\IEEEauthorblockA{\textsuperscript{2}
Instituto de Telecomunicacoes, Department of Electrical Engineering (DEE),\\
Universidade de Brasilia, Brasilia, Brazil}

\IEEEauthorblockA{\textsuperscript{3}
College of Science and Engineering, Hamad Bin Khalifa University, Doha, Qatar}

\IEEEauthorblockA{\textsuperscript{4}
Department of Electrical Engineering, University of Jordan, Amman, Jordan}

\IEEEauthorblockA{\textsuperscript{5}
Intelligent Information Processing Lab, Qatar University, Doha, Qatar}

\IEEEauthorblockA{
Email: kqaraqe@hbku.edu.qa
}

}

\maketitle
\thispagestyle{firstpagefooter}
\begin{abstract}
In this paper, we investigate the performance of physical-layer security of a pinching-antenna system on a lossless dielectric waveguide. In particular, the system uses a single pinching-antenna to convey confidential information from a base station to a legitimate destination equipped with a single antenna, while an eavesdropper, also equipped with a single antenna, attempts to decode the transmitted information. As such, the performance of the pinching-antenna system is evaluated in terms of average secrecy capacity, strictly positive secrecy capacity, and secrecy outage probability. To this end, accurate mathematical expressions for the aforementioned performance metrics are provided. To validate the analysis, the analytical results are numerically evaluated and further validated through Monte-Carlo simulations. The results demonstrate that secrecy capacity between the base station and the legitimate destination improves when the height of the pinching-antenna placed closer to the destination. Additionally, the performance can be improved when the eavesdropper's location over a rectangular area increases.

 \end{abstract}
 

\section{Introduction}

\IEEEPARstart{F}{lexible-antenna} systems are innovative techniques for dynamically adjusting wireless channel conditions, with pinching-antennas recently emerging as a novel evolution of this technology. The pinching-antenna is a versatile technology designed to establish robust line-of-sight links and mitigate large-scale path loss~\cite{Suzuki}. 
The mentioned technology is idealized by using small dielectric particles on a waveguide and presents the installation flexibility as its key feature~\cite{Ding}.

Over the years, the pinching-antenna systems have been studied in different contexts. Practical designs, as well as the performance analysis of pinching-antenna systems, are presented in~\cite{Ding}.
In~\cite{Yang}, its fundamentals, applications, challenges and key future research directions are described.
A study on antenna activation for non-orthogonal multiple access (NOMA)-assisted pinching-antenna systems is studied in~\cite{Wang}, where a matching-based algorithm is proposed to maximize the sum rate with low complexity.
The authors in~\cite{Xu} propose an algorithm to maximize the downlink data rate by optimizing  the locations of the pinching antennas.

The mentioned algorithm operates in two stages, the first of which consists of optimizing the locations of the pinching antennas to minimize the large-scale path loss, and the second aiming to refine the antenna locations to maximize the received signal strength.

Taking into account practical implementations, a new low-complexity placement design is also provided in~\cite{Xie}.
The optimization of the uplink performance of multiuser pinching-antenna systems is presented in~\cite{Tegos}, focusing on maximizing the minimum achievable data rate between devices. In \cite{tyrovolas2025}, the performance of outage probability and average rate is analyzed and evaluated under pinching-antenna systems serving a single user. Additionally, to maximize the system's performance of the system, the optimal location for the user is identified.

Recently, the performance of secrecy communications based on flexible-antenna (e.g., fluid antenna) systems has attracted the attention of researchers in the field. In \cite{10694739}, the authors investigated the performance of physical-layer security in fluid antenna systems under arbitrary correlated fading channels. Specifically, a transmitter equipped with a single fixed-antenna  aims to send confidential information to a legitimate destination equipped with a
planar fluid antenna system, while an eavesdropper, also equipped with a planar fluid antenna system, attempts to decode the transmitted information. In \cite{10468625}, the authors analyzed the performance of a fluid antenna communication system in terms of secrecy outage probability (SOP), considering that the fluid antenna system experiences spatially correlated Nakagami-$m$ fading. Additionally, the authors considered both non-diversity fluid antennas and a maximum-gain combining fluid antenna diversity scheme at the legitimate destination. Furthermore, multiple antennas performing maximal ratio combining at the eavesdropper are also considered. In \cite{10092780}, a physical layer security approach is proposed in which a base station transmits confidential information to a legitimate destination and an artificial noise signal to an eavesdropper. The legitimate destination is equipped with a fluid antenna to mitigate the impact of the artificial noise signal. To further enhance the secrecy rate, the authors proposed a power allocation scheme, assuming that channel state information (CSI) is available at the base station. In \cite{ghadi2024}, the authors investigated the influence of fluid antenna systems on secure communication networks assisted by reconfigurable intelligent surfaces (RIS). Specifically, the analysis focuses on a conventional wiretap channel scenario, wherein a transmitter with a fixed antenna array transmits confidential data to a legitimate destination equipped with fluid antenna system, facilitated by an RIS. Simultaneously, an eavesdropper, also equipped with fluid antenna system, seeks to intercept and decode the transmitted information. In \cite{10569014}, secret communication for fluid antenna systems is explored. Unlike conventional jamming schemes, where Gaussian noise interferes with both the legitimate destination and eavesdropper, the authors proposed sending encoded codewords to induce jamming at the eavesdropper. Recently, the work in \cite{sun2025} explored the joint design of baseband beamforming and pinching beamforming in a pinching-antenna systems-enabled multiuser wiretap channel. In the single-user scenario, a closed-form solution for the optimal baseband beamformer is derived, while a gradient-based optimization method is developed to determine the optimal placement of pinching antennas for maximizing the secrecy rate. For multiuser scenario, the authors introduced a fractional programming-based block coordinate descent algorithm, termed algorithm that integrates the Gauss–Seidel method with a one-dimensional search strategy to achieve efficient joint beamforming design.
  
In this paper, we analyze the performance of pinching-antenna systems based physical-layer security. Specifically, we derive expressions for key secrecy performance metrics, and to the best of the authors' knowledge, the analyses and all the expressions presented here are novels in the literature. Specifically, we derive expressions for the average secrecy capacity (SC), strictly positive secrecy capacity (SPSC), and SOP. These performance metrics are then used to evaluate the system's performance under different parameters.

The remainder of the paper is organized as follows. 
Section~\ref{sec2} describes the system model. 
Some important secrecy performance metrics namely the average SC, SPSC, and the SOP are presented in Sections~\ref{sec3}. Section~\ref{sec4} provides some numerical results. Finally, Section~\ref{sec5} brings the conclusions of the paper.

\section{System Model}\label{sec2}

\begin{figure}[t!] \centering
         \includegraphics[width=0.90\linewidth,keepaspectratio]{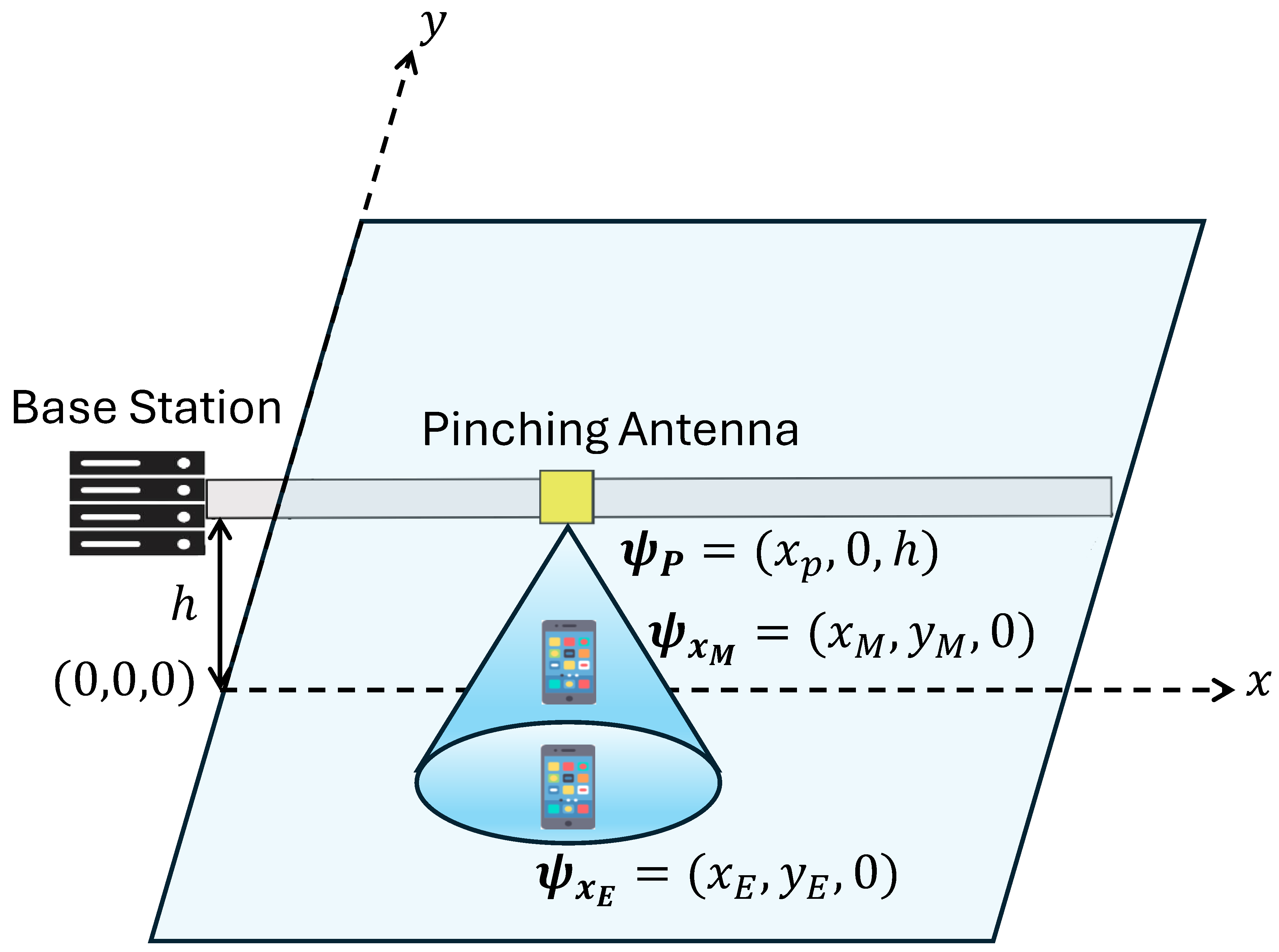}%
      \vspace{-3.0mm} \caption{A pinching-antenna system with a legitimate destination and an eavesdropper.}\label{sysm}
\end{figure}
Consider a downlink communication system, as shown in Fig. \ref{sysm}. In this system, a base station (BS) is located at $(0,0,0)$ and uses a single pinching-antenna based on a lossless dielectric waveguide parallel to the $x$-axis at a height $h$. Specifically, the pinching-antenna placed at $\boldsymbol{\psi_p} =(x_{p}, 0, h)$. The BS transmits confidential information to a main destination, $M$, which is randomly located within a rectangular area in the $x$-$y$ plane, with side lengths $D_{x_{M}}$ and $D_{y_{M}}$, respectively. The position of $M$ is denoted by $\boldsymbol{\psi_{x_{M}}} = (x_{M}, y_{M}, 0)$. An eavesdropper, located at $\boldsymbol{\psi_{x_{E}}} = (x_{E}, y_{E}, 0)$, attempts to intercept and decode the transmitted information. The locations of the main destination and the eavesdropper are randomly distributed. Specifically, $x_{i}$ is uniformly distributed over $[0, D_{x_{i}}]$, and $y_{i}$ is uniformly distributed over $\bigl[{-D_{y_{i}}\over2}, {D_{y_{i}}\over2}\bigr]$, where $i\in\{M, E\}$.    
 
Based on the system model of the wiretap channel \cite{Bloch}, the received signals at the main destination $M$ and the eavesdropper $E$ can be expressed, respectively, as
\begin{align} 
z_{M} =& \sqrt {P_{t}}\, h_{M}\, h_{p}\, s + n_{M}\\[-1pt]
z_{E} = &\sqrt {P_{t}\,}h_{E}\, h_{p}\, s + n_{E},
\end{align}
where $P_{t}$ is the transmit power, $s$ is the transmitted signal with $\mathbb{E}[|s|^{2}]=1$, $n_{M}$ and $n_{E}$ are zero-mean additive white Gaussian noise with variance of $\sigma^{2}$ at the main destination and eavesdropper, respectively, $h_{i}$ is the channel between
a pinching-antenna and the destination $i$, $h_{p}$ is the phase shift influence the signal transmitted by the pinching-antenna, and are expressed respectively as
\begin{align}
  h_{{i}} = \frac{\sqrt{\eta}}{\|\boldsymbol{\psi_x}_{i} -\boldsymbol{\psi_p}\|}\,\exp{\left(-j \frac{2\pi}{\lambda} \|\boldsymbol{\psi_x}_{i} -\boldsymbol{\psi_p}\|\right)} \end{align}
  and 
  \begin{align}
    h_{p} =  \exp{\left(-j \frac{2\pi}{\lambda_g} \|\boldsymbol{\psi_p} -\boldsymbol{\psi_0}\|\right)},
\end{align}
where $\eta= \frac{\lambda^2}{16 \pi^2}$ is the path loss at a reference distance of $1$ m, $\lambda={c\over f_{}o}$ is the free-space wavelength, with $c$ being the speed of light and $f_{o}$ is the carrier frequency, $j=\sqrt{-1}$, and $\|\cdot\|$ is the Euclidean norm, $\lambda_g = \frac{\lambda}{n_{\mathrm{eff}}}$ is the guided wavelength, with  $n_{\mathrm{eff}}$ being the effective refractive index, and $\boldsymbol{\psi_0}= (0, 0, h)$ is the position of the waveguide feeding point. 

\textit{Corollary:} The PDF at either the main destination $M$ or the eavesdropper $E$ in a system with a single pinching-antenna on a lossless dielectric waveguide can be expressed as
\begin{align}\label{pdfi}
f_{\Gamma_i}(\gamma_{i}) =
\begin{cases}
\frac{1}{D_{y_i}} \frac{\eta \overline{\gamma}_{i}}{\gamma_i^2  \sqrt{\frac{\eta \overline{\gamma}_{i}}{\gamma_i } - h^2}}, & \frac{\eta \overline{\gamma}_{i}}{h^2 + \frac{D_{y_i}^2}{4} } \leq \gamma_i \leq \frac{\eta \overline{\gamma}_{i}}{h^2 }, \\[10pt]
0, & \text{otherwise},
\end{cases}
\end{align}
where $\overline{\gamma}_{i} = {P_{t}\over\sigma^{2}}$ is the average signal-to-noise ratio (SNR).
\begin{IEEEproof}
With the help of \cite[Proposition 2]{tyrovolas2025}, the CDF at either the main destination $M$ or the eavesdropper $E$ can be expressed as
\begin{align}\label{cdfi}
F_{\Gamma_i}(\gamma_{i}) =
\begin{cases}
1, & \gamma_i \geq \frac{\eta \overline{\gamma}_{i}}{h^2 }, \\[10pt]
1 - \frac{2}{D_{y_i}} \sqrt{\frac{\eta \overline{\gamma}_{i}}{\gamma_i } - h^2}, & \frac{\eta \overline{\gamma}_{i}}{h^2 + \frac{D_{y_i}^2}{4} } \leq \gamma_i \leq \frac{\eta \overline{\gamma}_{i}}{h^2 }, \\[10pt]
0, & \gamma_i \leq \frac{\eta \overline{\gamma}_{i}}{h^2 + \frac{D_{y_i}^2}{4} }.
\end{cases}
\end{align}
Capitalizing on the fact $f_{\Gamma_i}(\gamma_{i})={d\over d\gamma_{i}}F_{\Gamma_i}(\gamma_{i})$, the PDF in \eqref{pdfi} is obtained.
\end{IEEEproof}

\section{Secrecy Analysis}\label{sec3}
\subsection{Average Secrecy Capacity}
The average SC \( \mathcal{C}_s^{\text{ASC}} \) is defined as the expected value of the secrecy capacity \( \mathcal{C}_s\), which is the maximum achievable secrecy rate given that the channel state information of the eavesdropper's channel in known at the transmitting BS. Therefore, \( \mathcal{C}_s\) can be defined as \cite{Bloch}:
\begin{align}
\mathcal{C}_s =
\begin{cases}
\log_{2} (1 + \gamma_M) - \log_{2} (1 + \gamma_E), & \gamma_M > \gamma_E \\
0, & \gamma_M \leq \gamma_E,
\end{cases}
\end{align}
where \( \gamma_M \) denotes the SNR at the main (legitimate) destination and \( \gamma_E \) denotes is the SNR at the eavesdropper. Thus, the average SC (in bps/Hz) is computed as:
\begin{align}\label{secrecy}
\mathcal{C}_s^{\text{ASC}} = \mathbb{E}[\mathcal{C}_s] = \int\limits_0^\infty \int\limits_0^\infty \mathcal{C}_s\, f_{\Gamma_M, \Gamma_E}(\gamma_M, \gamma_E) \, {\mathrm d}\gamma_M \, {\mathrm d}\gamma_E,
\end{align}
where \(f_{\Gamma_M, \Gamma_E}(\gamma_M, \gamma_E)\) is the joint PDF of \( \Gamma_M \) and \( \Gamma_E \). 

Since \( \mathcal{C}_s\) is nonzero only when \( \gamma_M > \gamma_E \) and given that \( \gamma_M \) and \( \gamma_E \) are independent, the integral in \eqref{secrecy} simplifies to
\begin{align}\label{secrecy1}
&\mathcal{C}_s^{\text{ASC}} = {1\over\ln{(2)}} \int\limits_0^\infty f_{\Gamma_M}(\gamma_M) \cr&\times\!
 \left( \int\limits_0^{\gamma_M} \bigg( \ln (1 + \gamma_M) - \ln (1 + \gamma_E) \bigg) f_{\Gamma_E}(\gamma_E) \, {\mathrm d}\gamma_E \right)\! {\mathrm d}\gamma_M.&
\end{align}

Knowing that: (i) the PDFs \( f_{\Gamma_M}(\gamma_M) \) and \( f_{\Gamma_E}(\gamma_E) \) are nonzero only in specific ranges, i.e., \(
  \frac{\eta \overline{\gamma}_{M}}{h^2 + \frac{D_{y_M}^2}{4}} \leq \gamma_M \leq \frac{\eta \overline{\gamma}_{M}}{h^2},
  \) for \( \gamma_M \), \(\frac{\eta \overline{\gamma}_{E}}{h^2 + \frac{D_{y_E}^2}{4}} \leq \gamma_E \leq \frac{\eta \overline{\gamma}_{E}}{h^2}\) for \( \gamma_E \), and (ii) the inner integral requires \( \gamma_E \leq \gamma_M \), thus the upper limit for \( \gamma_E \) is \( \gamma_M \). Therefore, the integral in \eqref{secrecy1} becomes
\begin{align}\label{finASC}
&\mathcal{C}_s^{\text{ASC}} = {1\over\ln{(2)}}\int\limits_{\frac{\eta \overline{\gamma}_{M}}{h^2 + \frac{D_{y_M}^2}{4}}}^{\frac{\eta \overline{\gamma}_{M}}{h^2}} f_{\Gamma_M}(\gamma_M) \cr&\left(\!\! \int\limits_{\frac{\eta \overline{\gamma}_{E}}{h^2 + \frac{D_{y_E}^2}{4}}}^{\min(\gamma_M, \frac{\eta \overline{\gamma}_{E}}{h^2})} \!\bigg(\! \ln(1 + \gamma_M) - \ln(1 + \gamma_E)\! \bigg) f_{\Gamma_E}(\gamma_E) \, {\mathrm d}\gamma_E \!\!\right)\! {\mathrm d}\gamma_M.\cr&
\end{align}
By substituting the PDFs of \( f_{\Gamma_M}(\gamma_M) \) and \( f_{\Gamma_E}(\gamma_E) \) from \eqref{pdfi}, the average SC can be evaluated as
\begin{align}\label{finASC1}
&\mathcal{C}_s^{\text{ASC}} = {\eta^{2}\, \overline{\gamma}_{M}\,\overline{\gamma}_{E}\over D_{y_M} D_{y_E}\ln{(2)}}\int\limits_{\frac{\eta \overline{\gamma}_{M}}{h^2 + \frac{D_{y_M}^2}{4}}}^{\frac{\eta \overline{\gamma}_{M}}{h^2}}  \frac{1}{\gamma_M^2  \sqrt{\frac{\eta \overline{\gamma}_{M}}{\gamma_M } - h^2}} \cr&\times\left(\!\! \int\limits_{\frac{\eta \overline{\gamma}_{E}}{h^2 + \frac{D_{y_E}^2}{4}}}^{\min(\gamma_M, \frac{\eta \overline{\gamma}_{E}}{h^2})}\,\,\, \frac{ \ln(1 + \gamma_M) - \ln(1 + \gamma_E)}{\gamma_E^2  \sqrt{\frac{\eta \overline{\gamma}_{E}}{\gamma_E } - h^2}} \, {\mathrm d}\gamma_E \!\!\right)\! {\mathrm d}\gamma_M.\cr&
\end{align}

 \subsection{Strictly Positive Secrecy Capacity}
The existence of strictly positive secrecy capacity, between the transmitting BS and the main destination $M$, is determines by the SPSC. That is
\begin{align}
\mathcal{P}^{\text{SPSC}} &= \mathbb{P}\left( \mathcal{C}_{s} > 0 \right) = \mathbb{P}\left( \Gamma_{M} > \Gamma_{E} \right) \nonumber \\
&= \int_{0}^{\infty} f_{\Gamma_{M}}(\gamma_{M}) \, F_{\Gamma_{E}}(\gamma_{M}) \, {\mathrm d}\gamma_M.
\end{align}

Since the PDF \(f_{\Gamma_{M}}(\gamma_{M})\) is nonzero only within a specific range, the integral simplifies to
\begin{align}\label{spsc}
\mathcal{P}^{\text{SPSC}}=\int_{\frac{\eta \overline{\gamma}_{M}}{h^2 + \frac{D_{y_M}^2}{4}}}^{\frac{\eta \overline{\gamma}_{M}}{h^2}} \left( \frac{1}{D_{y_M}}  \frac{\eta \overline{\gamma}_{M}}{\gamma_{M}^{2} \sqrt{\frac{\eta \overline{\gamma}_{M}}{\gamma_{M}} - h^2}} \right)  F_{\Gamma_E}(\gamma_M) \, {\mathrm d}\gamma_M.
\end{align}
The CDF \( F_{\Gamma_E}(\gamma_M) \) has three cases, which overlap with the range of \( \gamma_M \) defined by \( f_{\gamma_M}(\gamma_M) \). That is,

\subsubsection*{Case 1: \( \gamma_M \geq \frac{\eta \overline{\gamma}_{E}}{h^2} \)}

In this case, \( F_{\gamma_E}(\gamma_M) = 1 \). This range overlaps with \( f_{\gamma_M}(\gamma_M) \) only if:
\[
\frac{\eta \overline{\gamma}_{M}}{h^2 + \frac{D_{y_M}^2}{4}} \leq \gamma_M \leq \frac{\eta \overline{\gamma}_{M}}{h^2} \quad \text{and} \quad \gamma_M \geq \frac{\eta \overline{\gamma}_{E}}{h^2}.
\]
Thus, the lower limit for this case is:
\[
\gamma_{{L_{1}}} = \max\left( \frac{\eta \overline{\gamma}_{M}}{h^2 + \frac{D_{y_M}^2}{4}}, \frac{\eta \overline{\gamma}_{E}}{h^2} \right).
\]
While the upper limit is:
\[
\gamma_{{U_{1}}} = \frac{\eta \overline{\gamma}_{M}}{h^2}.
\]

\subsubsection*{Case 2: \( \frac{\eta \overline{\gamma}_{E}}{h^2 + \frac{D_{y_E}^2}{4}} \leq \gamma_M \leq \frac{\eta \overline{\gamma}_{E}}{h^2} \)}

In this case, \( F_{\gamma_E}(\gamma_M) = 1 - \frac{2}{D_{y_E}} \sqrt{\frac{\eta \overline{\gamma}_{E}}{\gamma_M} - h^2} \). This range overlaps with \( f_{\gamma_M}(\gamma_M) \) only if:
\[
\frac{\eta \overline{\gamma}_{M}}{h^2 + \frac{D_{y_M}^2}{4}} \leq \gamma_M \leq \frac{\eta \overline{\gamma}_{M}}{h^2} \quad \text{and} \quad \frac{\eta \overline{\gamma}_{E}}{h^2 + \frac{D_{y_E}^2}{4}} \leq \gamma_M \leq \frac{\eta \overline{\gamma}_{E}}{h^2}.
\]
Therefore, the lower limit for this case is:
\[
\gamma_{{L_{2}}} = \max\left( \frac{\eta \overline{\gamma}_{M}}{h^2 + \frac{D_{y_M}^2}{4}}, \frac{\eta \overline{\gamma}_{E}}{h^2 + \frac{D_{y_E}^2}{4}} \right).
\]
While the upper limit is:
\[
\gamma_{{U_{2}}} = \min\left( \frac{\eta \overline{\gamma}_{M}}{h^2}, \frac{\eta \overline{\gamma}_{E}}{h^2} \right).
\]

\subsubsection*{Case 3: \( \gamma_M \leq \frac{\eta \overline{\gamma}_{E}}{h^2 + \frac{D_{y_E}^2}{4}} \)}

In this case, \( F_{\gamma_E}(\gamma) = 0 \).

Based on the above cases, the integral in \eqref{spsc} can be rewritten as:
\begin{align}\label{solint}
&\mathcal{P}^{\text{SPSC}}= \frac{2\eta \overline{\gamma}_{M}}{D_{y_M}}\int_{\gamma_{{L_{1}}}}^{\gamma_{{U_{1}}}}  \frac{1}{\gamma_M^2  \sqrt{\frac{\eta \overline{\gamma}_{M}}{\gamma_M } - h^2}}\, {\mathrm d}\gamma_M
 \cr& - \frac{2\eta \overline{\gamma}_{M}}{D_{y_M} D_{y_E}}\int_{\gamma_{{L_{2}}}}^{\gamma_{{U_{2}}}} \frac{\sqrt{\frac{\eta \overline{\gamma}_{E}}{\gamma_M} - h^2}}{\gamma_M^2  \sqrt{\frac{\eta \overline{\gamma}_{M}}{\gamma_M } - h^2}} \,  {\mathrm d}\gamma_M.&
\end{align}
Finally, the SPSC can be evaluated as:
\begin{align}\label{solint1}
&\mathcal{P}^{\text{SPSC}}= F_{\Gamma_{M}}(\gamma_{{U_{1}}}) - F_{\Gamma_{M}}(\gamma_{{L_{1}}})
 \cr& - \frac{2\eta \overline{\gamma}_{M}}{D_{y_M} D_{y_E}}\int_{\gamma_{{L_{2}}}}^{\gamma_{{U_{2}}}} \frac{\sqrt{\frac{\eta \overline{\gamma}_{E}}{\gamma_M} - h^2}}{\gamma_M^2  \sqrt{\frac{\eta \overline{\gamma}_{M}}{\gamma_M } - h^2}} \,  {\mathrm d}\gamma_M.&
\end{align}
\subsection{Secrecy Outage Probability}
Here, we consider the case when the transmitting BS and main destination $M$ have no CSI knowledge about
the eavesdropper (i.e., also known as a passive eavesdropper). In this case, the SOP can be used as a key performance metric to characterize the main link as it denotes the probability that the instantaneous secrecy capacity falls below a target secrecy rate $R_{s}$. Thus, The SOP is given by \cite{Bloch} :
\begin{align}\label{sop}
\mathcal{P}^{\text{SOP}} = \int _{0}^\infty F_{\gamma_M}\left (\Psi \,{\gamma _{E}} + \Psi - 1 \right) f_{\gamma_E}\left (\gamma _{E} \right) {\mathrm d}\gamma_E,
\end{align}
where $\Psi = \exp{(R_s)}\geq 1$, with $R_s$  (in bps/Hz) being the target secrecy rate.
Since the PDF \(f_{\Gamma_{E}}(\gamma_{E})\) is nonzero only within a specific range, the integral simplifies to
\begin{align}\label{sop1}
\mathcal{P}^{\text{SOP}} = \int_{\frac{\eta \overline{\gamma}_E}{h^2 + \frac{D_{y_E}^2}{4}}}^{\frac{\eta \overline{\gamma}_E}{h^2}} F_{\Gamma_M}\left (\Psi \,{\gamma _{E}} + \Psi - 1 \right) f_{\Gamma_E}\left (\gamma _{E} \right) {\mathrm d}\gamma_E.
\end{align}

Based on $F_{\Gamma_M}(\cdot)$ and $f_{\Gamma_E}(\cdot)$, respectively given by \eqref{cdfi} and \eqref{pdfi}, the latter integral can be rewritten as
\begin{align}\label{sop2}
&\mathcal{P}^{\text{SOP}} = 1 - \int_{\frac{\eta \overline{\gamma}_E}{h^2 + \frac{D_{y_E}^2}{4}}}^{\frac{\eta \overline{\gamma}_E}{h^2}} f_{\Gamma_E}\left(\gamma _{E}\right) {\mathrm d}\gamma_E 
\cr&\!\!+\!\int_{\frac{\eta \overline{\gamma}_E}{h^2 + \frac{D_{y_E}^2}{4}}}^{\frac{\eta \overline{\gamma}_E}{h^2}}\!\! \left(\frac{2}{D_{y_M}} \sqrt{\frac{\eta \overline{\gamma}_{M}}{\Psi \gamma_E + \Psi - 1} - h^2}\right) f_{\Gamma_E}\left (\gamma _{E}\right) {\mathrm d}\gamma_E.\,\,\,\,&
\end{align}
The first integral, based on the definition of the total probability of a given PDF, is equal to one. Thus, the SOP can be evaluated as:  
\begin{align}\label{sop3}
\mathcal{P}^{\text{SOP}} = \frac{2\eta \overline{\gamma}_{E}}{D_{y_M} D_{y_E}}\int_{\frac{\eta \overline{\gamma}_E}{h^2 + \frac{D_{y_E}^2}{4}}}^{\frac{\eta \overline{\gamma}_E}{h^2}}
     \frac{\sqrt{\frac{\eta \overline{\gamma}_{M}}{\Psi \gamma_E + \Psi - 1} - h^2}}{\gamma_E^2  \sqrt{\frac{\eta \overline{\gamma}_{E}}{\gamma_E } - h^2}} \,\, {\mathrm d}\gamma_E.
\end{align}
\section{Numerical Results}\label{sec4}
In this section, we evaluate the performance of physical-layer security for the pinching-antenna system. The noise power is set to $\sigma^{2}=-90$ dBm and the carrier frequency is $f_{o}=28$ GHz. The dimensions of the area are set to $D_{x_{M}}\times D_{y_{M}} =10\times 10\, \text{m}^{2}$ for the main destination and to $D_{x_{E}}\times D_{y_{E}} =10\times 10\, \text{m}^{2}$ for the eavesdropper, unless otherwise stated. The results for the average SC,  SPSC, and SOP are numerically obtained using \eqref{finASC1}, \eqref{solint}, and \eqref{sop3}, respectively. The numerical results are further validated through Monte-Carlo simulations, and both sets of results are in excellent agreement.

\begin{figure}[t!] \centering
         \includegraphics[width=\figwidth,keepaspectratio]{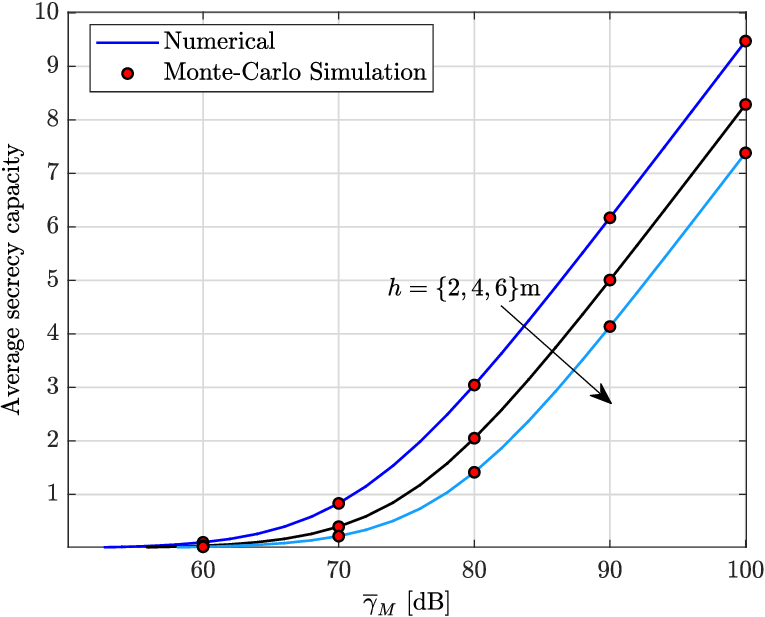}%
      \vspace{-3.0mm} \caption{Performance of average SC for the pinching-antenna system. $\overline{\gamma}_{E}=50$ dB.}\label{fig1}
\end{figure}

Fig. \ref{fig1} shows the results for the average secrecy capacity for different heights, i.e., $h=\{2, 4, 6\}$ m, of the dielectric waveguide (i.e., the height of the pinching-antenna) with $\overline{\gamma}_{E}=50$ dB. The results demonstrate that the average secrecy capacity improves as the height decreases. This is because, as the height decreases, the pinching-antenna becomes closer to the destination, thereby improving performance. For example, at $\overline{\gamma}_{M}=80$ dB and when $h$ decreases from $6$ m to $2$, the average SC increase from $1.5$ to $3$ [bps/Hz], showing a significant improvement of $100\%$. Additionally, the average SC improves as the average received SNR at the main destination, $\overline{\gamma}_{M}$, increases. 

\begin{figure}[t!] \centering
         \includegraphics[width=\figwidth,keepaspectratio]{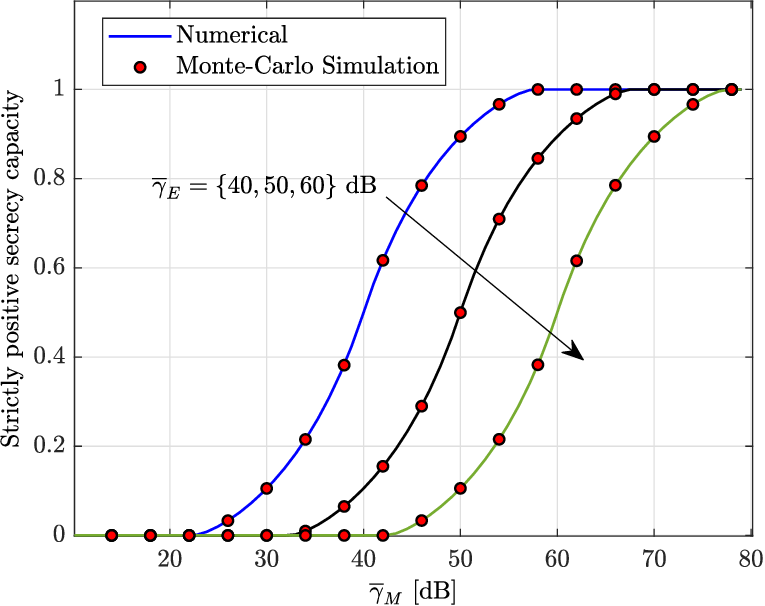}%
      \vspace{-3.0mm} \caption{Performance of SPSC for PA system. $h=2$ m, $D_{x_{M}} = D_{x_{E}}=10$ m, and $D_{y_{M}} = D_{y_{E}}=30$ m.}\label{fig2}
\end{figure}

The performance of SPSC for different values of the average SNR, $\overline{\gamma}_{E}$, at the eavesdropper with $h=2$ m, $D_{x_{M}} = D_{x_{E}}=10$ m, and $D_{y_{M}} = D_{y_{E}}=30$ m is depicted in Fig. \ref{fig2}. From the results, it can be seen that as $\overline{\gamma}_{E}$ decreases, the probability of the existence of positive secrecy capacity increases, indicating better performance. For example, at $\overline{\gamma}_{M}=50$ dB and when $\overline{\gamma}_{E}$ decreases from $50$ dB to $40$ dB, The value of the SPSC increases from $0.5$ to $0.9$, showing a significant improvement of $80\%$.

\begin{figure}[t!] \centering
         \includegraphics[width=\figwidth,keepaspectratio]{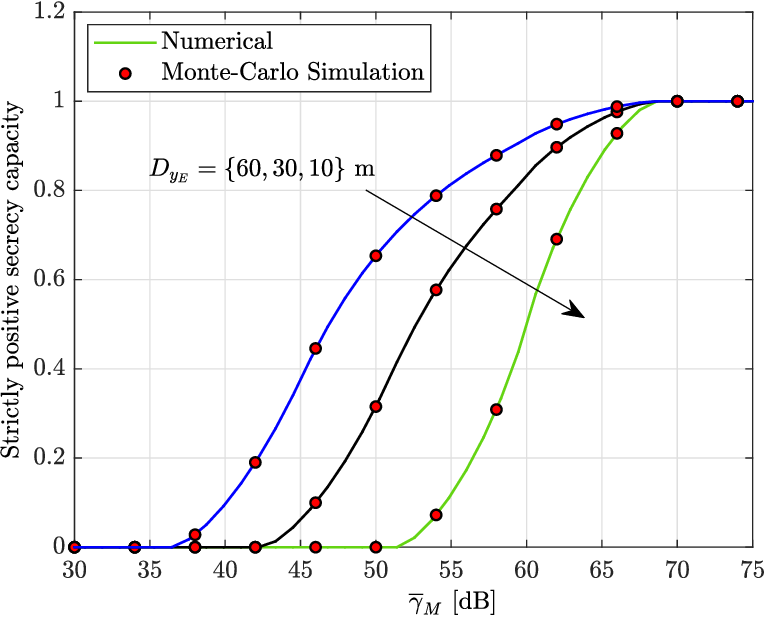}%
      \vspace{-3.0mm} \caption{Performance of SPSC for the pinching-antenna system. $\overline{\gamma}_{E}=60$ dB, $h=2$ m,  $D_{x_{M}} =D_{x_{E}}=D_{y_{M}}=10$ m.}\label{fig3}
\end{figure}

Fig. \ref{fig3} illustrates the performance of SPSC for different values of the dimension of the eavesdropper, i.e., $D_{y_{E}}=\{60, 30, 10\}$ m. The results demonstrate that as $D_{y_{E}}$ decreases, the SPSC improves. This is because, as $D_{y_{E}}$ decreases, the probability of the eavesdropper's location being closer to the pinching-antenna increases, resulting in smaller path losses. For instance,  at $\overline{\gamma}_{M}=50$ dB and when $D_{y_{E}}$ decreases from $60$ m to $30$ m, the value of the SPSC increases from $0.32$ to $0.65$, showing a significant improvement of $103\%$. Furthermore, we have observed that, under the conditions where $\overline{\gamma}_{E}=\overline{\gamma}_{M}$, $D_{x_{E}}=D_{x_{M}}$, and $D_{y_{E}}=D_{y_{M}}$, the probability of SPSC is equal to $0.5$. As  $\overline{\gamma}_{M}$ begins to exceed $\overline{\gamma}_{E}$, the probability of SPSC increases, indicating improved performance. 
  
\begin{figure}[t!] \centering
         \includegraphics[width=\figwidth,keepaspectratio]{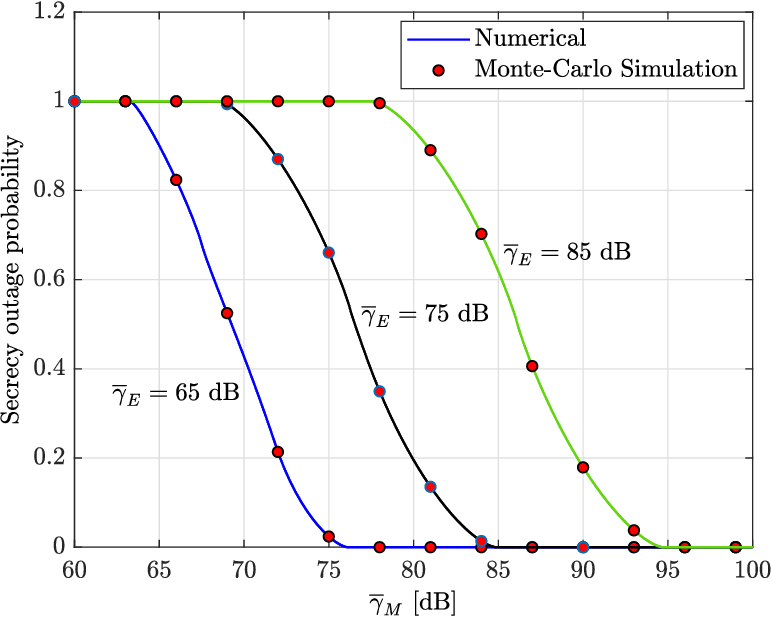}%
      \vspace{-3.0mm} \caption{Performance of SOP for the pinching-antenna system. $R_{s} = 0.25$ bps/Hz and $h=2$ m.}\label{fig4}
\end{figure}

The performance of SOP under varying the average SNR, $\overline{\gamma}_{E}=\{65, 75, 85\}$ dB, at the eavesdropper for $h=2$ m and $R_s=0.25$ bps/Hz is plotted in Fig. \ref{fig4}. The results show that as $\overline{\gamma}_{M}$ increases, the value of the SOP decreases, indicating better performance. On the other hand, the results show that as $\overline{\gamma}_{E}$ increases, the value of the SOP increases, indicating worse performance.

\section{Conclusion}\label{sec5}
In this paper, we have analyzed the performance of pinching-antenna system in the context of physical-layer security. In particular, mathematical expressions for the average SC, SPSC, and SOP were derived. Numerical results were obtained for some representative scenarios and the further validated through Monte-Carlo simulation results. The results demonstrated that the average SC performance improves as the height of the dielectric waveguide, i.e., the height of the pinching-antenna, decreases. Additionally, the results showed that the probability of the existence of positive secrecy capacity, i.e.,  the SPSC, increases as the dimension of the eavesdropper's location increases, i.e., $D_{y_{E}}$ and(or) the average received SNR of the eavesdropper $\overline{\gamma}_{E}$ decreases. Moreover, the results illustrated that the SOP improves as the average received SNR of the main destination $\overline{\gamma}_{M}$ increases and(or) $\overline{\gamma}_{E}$ decreases.

\section*{Acknowledgment}

The research with Tamer M. Khattab is supported by the Qatar Research Development and Innovation (QRDI) through QNRF under Grant AICC03-0530-200033.

This work was supported by the Qatar Research Development and Innovation Council (QRDI) under Grant No. NPRP14C-0909-210008 and by research funding from Hamad Bin Khalifa University under the Thematic Research Grant Program Cycle 3. The statements made herein are solely the responsibility of the authors. The content is solely the responsibility of the authors and does not necessarily represent the official views of QRDI.

\bibliographystyle{IEEEtran}
\bibliography{akefRefe}

\end{document}